\documentclass{iaus}
\usepackage{graphicx}

\title[Seyfert 2 Nuclei] %% give here short title %%
{The Stellar Populations of Seyfert 2 Nuclei}

\author[Sarzi, Shields, Pogge, Martini]   %% give here short author list %%
{Marc Sarzi$^1$, Joseph C. Shields$^2$, Richard W. Pogge,  \break \and Paul Martini$^3$}

\affiliation{
$^1$Centre for Astrophysics Research, University
of Hertfordshire, AL10 9AB Hatfield, UK\\

$^2$Physics \& Astronomy Department, Ohio
University, Athens, OH 45701 USA\\

$^3$Department of Astronomy, The Ohio
State University, Columbus, OH 43210, USA
}

\pubyear{2004}
\volume{xxx}  %% insert here IAU Symposium No.
\pagerange{119--126}
\date{?? and in revised form ??}
\jname{Proceedings Title IAU Symposium}
\editors{A.C. Editor, B.D. Editor \& C.E. Editor, eds.}

\newcommand\kms{$\rm km\,s^{-1}$}

\newcommand{\Ha}{H$\alpha$}

\newcommand{\Nii}{[{\sc N$\,$ii}]}

\newcommand{\Nev}{[{\sc N$\rm e$v}]}
\newcommand{\Feiii}{[{\sc F$\rm e$iii}]}
\newcommand{\Ariv}{[{\sc A$\rm r$iv}]}
\newcommand{\Arv}{[{\sc A$\rm r$v}]}

\newcommand{\HST}{{\it HST\/}}
\newcommand{\placefigone}{
\begin{figure}
\begin{center}
\includegraphics[width=0.9\textwidth, trim = 15 50 13 0]{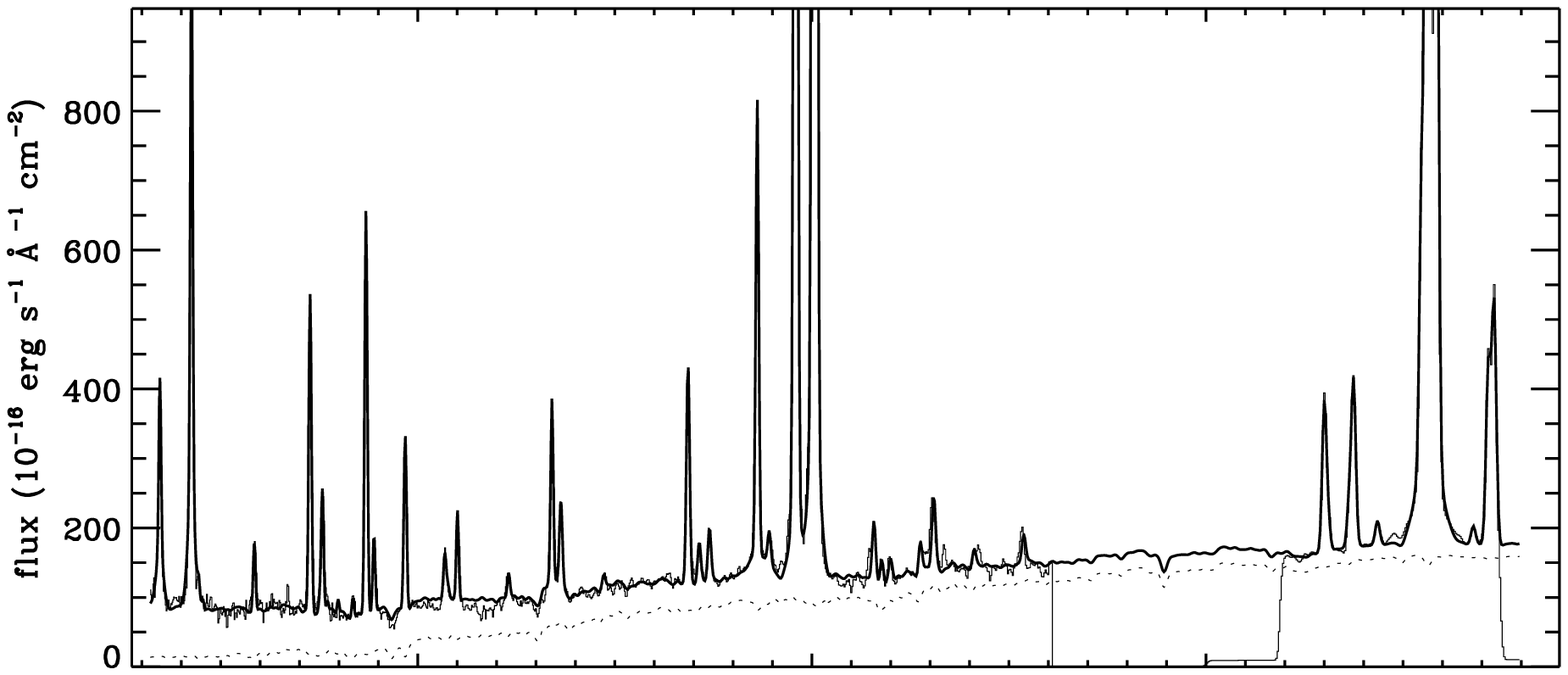}
\includegraphics[width=0.9\textwidth, trim = 15 10 13 0]{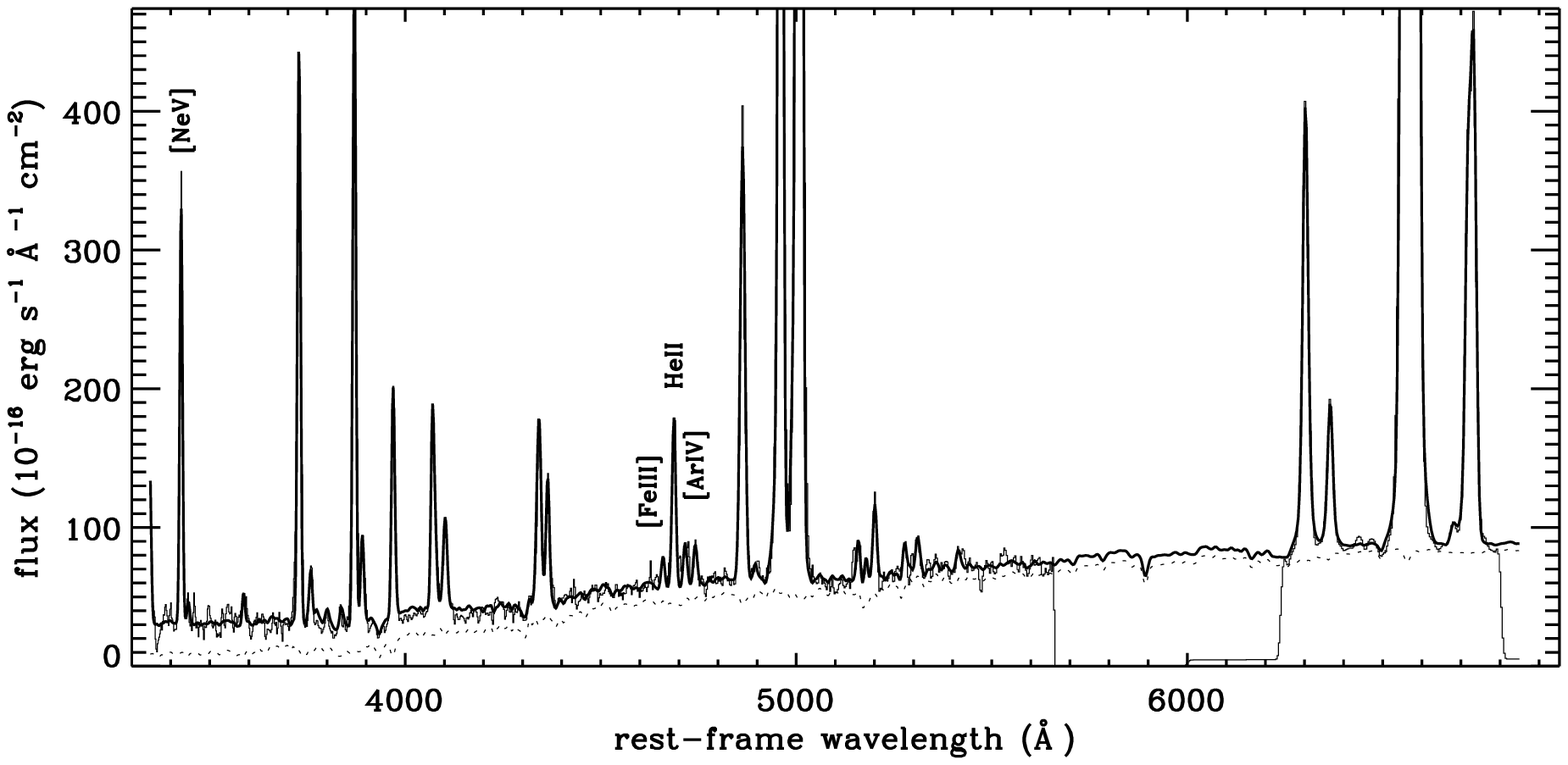}
\end{center}
\caption{Combined G430L and G750M spectrum for the central
$0''\!.25\times0''\!.2$ of the Seyfert2 nuclei in Mrk573 (up) and
NGC5427 (down). Overplotted to the data is the best fitting
combination of single-age stellar population models from Bruzual \&
Charlot (2003) and emission-line Gaussian templates. The dotted line
shows the contribution of the oldest, 10-Gyr-old, stellar template. A
blue continuum is needed to match the spectrum of these nuclei.}
\label{fig:fit}
\end{figure}
}

\newcommand{\placefigtwo}{
\begin{figure}
\begin{center}
\includegraphics[width=0.9\textwidth, trim = 18 10 13 0]{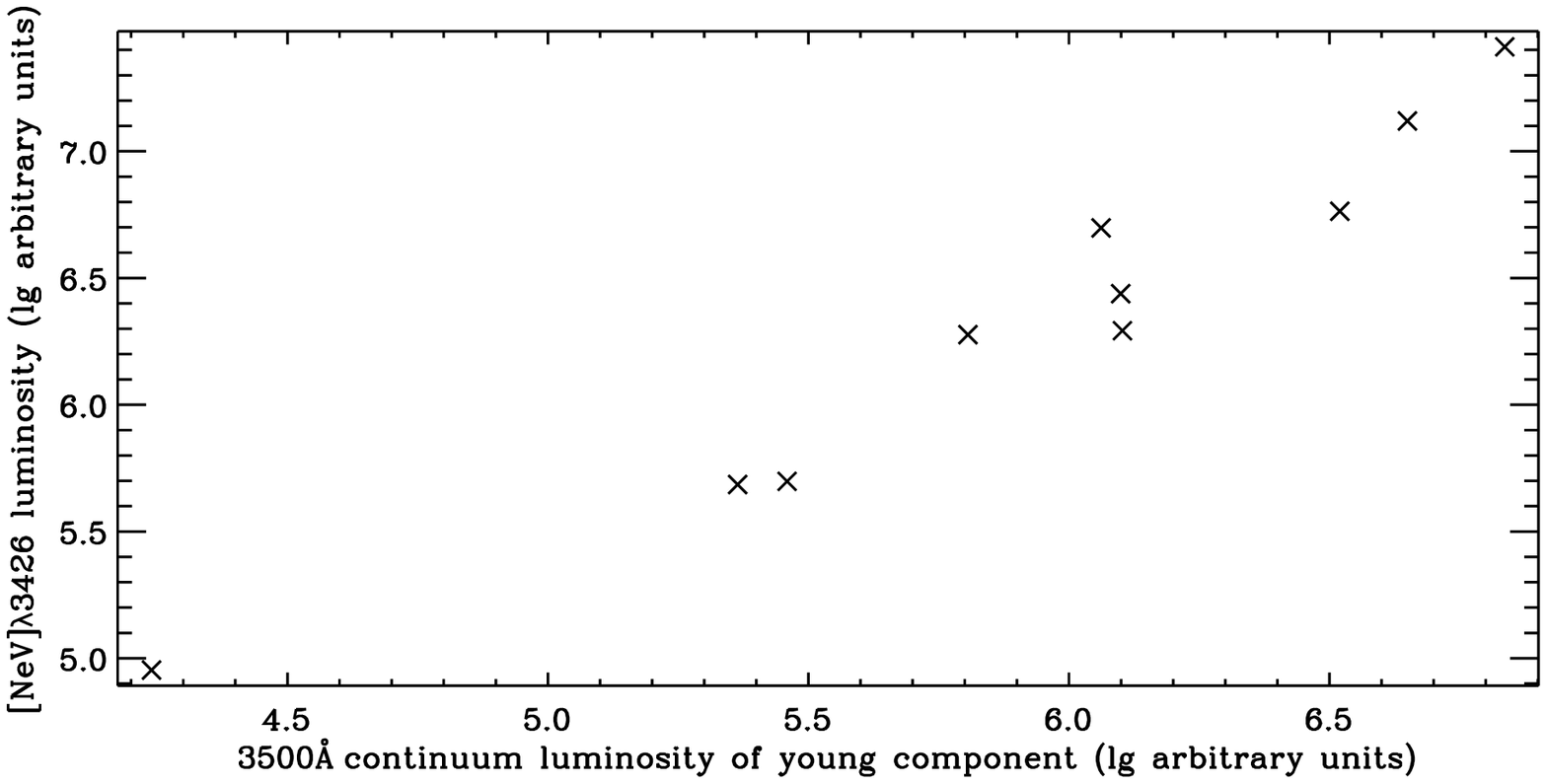}
\end{center}
\caption{Correlation between the luminosity of the featureless blue
continuum, obtained from rescaling the contribution to the observed
flux at 3500\AA\ of the youngest stellar template, and the strength of
the \Nev$\lambda$3426 nebular emission in Seyfert 2 nuclei.
}
\label{fig:Baldwin}
\end{figure}
}

\begin{document}

\maketitle

\begin{abstract}
We present a preliminary analysis of the stellar populations in the
central parsecs of a sample of 22 Seyfert 2 galaxies, based on a
careful separation of nebular emission and stellar light in
high-spatial resolution \HST-STIS spectra. 10\% of the surveyed nuclei
display stellar populations of intermediate age, $\sim$1-2 Gyr old,
whereas the remaining targets appear to be evenly split between
objects showing only very old stellar populations and nuclei requiring
also an additional blue featureless component, which we characterise
by means of very young, few-Myr-old stars.
The small fraction of stellar population of intermediate age seems to
argue against the presence of such a young component, however, since
the short lifetime of O-stars would imply recurrent star-formation
episodes and the build-up over the last 1-2 Gyr of a detectable
intermediate-age population.
Additionally, a correlation between the luminosity of such a blue
component and the emission from highly-ionised species, together with
the general absence of Wolf-Rayet features, further suggests that the
featureless continuum arises from the central engine rather than from
star-forming regions.
We discuss our results in the framework of the unification paradigm
and of models for star formation close to supermassive black holes.

\end{abstract}

\firstsection % if your document starts with a section,
\section{Introduction}

The study of the stellar population content of galactic nuclei is
crucial to our understanding of the AGN phenomenon, as it allows to
assess the relative r\^ole of star formation and accretion onto the
central supermassive black hole (SMBH) in powering the observed
nebular activity. In the specific case of Seyfert 2 nuclei (Sy2s),
constraining the presence of young stars is also relevant to
unification theory (Antonucci 1993).
While Sy2s might not show a broad-line region (BLR) due to the
unfavorable orientation of an obscuring torus, they nonetheless
exhibit a featureless blue continuum like that in their Type 1
counterparts. This would appear to be problematic for the unification
theory since such a component, which presumably arises from an
accretion disk, should also be blocked by the obscuring material. As a
possible solution, it has been suggested that the blue continuum of
Sy2s originates instead in star forming regions (e.g., Colina et
al. 1997; Gonz{\'a}lez Delgado et al. 1998). Yet, given the limited
spatial resolution characterising most spectroscopic studies of Sy2s,
typically subtending a few 100 pc, it is still unclear whether such
star formation is truly nuclear.

To address this issue we have carried out an observing campaign of
nearby Sy2s with the STIS spectrograph on board the Hubble Space
Telescope (\HST), in order to constrain the amount of star formation
as close as possible to the center.

\section{Sample and Data} 

Our sample consists of 19 nearby Seyfert 2 galaxies that were selected
from the combined CfA and Revised Shapley-Ames Seyfert samples
(Osterbrock \& Martel 1993; Mulchaey et al. 1997) to have receding
velocities $v<5000$ \kms and a distinct, unobscured nucleus from
archival \HST\ images.
To this sample we added 3 Seyfert 2 galaxies from the Survey of Nearby
Nuclei with STIS (SUNNS, Shields et al. 2007), which in turn were
selected from the Palomar spectroscopic survey (Ho et al. 1997) based
only on proximity ($D<17$ Mpc) and a minimum flux of \Ha+\Nii\ line
emission.

For all these Sy2s, two-dimensional spectra were obtained with
the G430L and G750M gratings, yielding spectra spanning $3300-5700$
and $6300-6850$\AA\ with FWHM spectral resolution for extended sources
of 10.9 and 2.2\AA, respectively. From these data we extracted
one-dimensional spectra over a $0''\!.25\times0''\!.2$ aperture,
corresponding to a circular aperture with a radius $R=0''\!.13$, or 29
pc for the mean sample distance of $\sim46$ Mpc.

\section{Method}   

In order to place constraints on the star-formation history of our
sample Sy2s, we modeled the stellar continuum in the STIS spectra with
linear combinations of single-age stellar population models from
Bruzual \& Charlot (2003), with weights and velocity broadening
optimized using the direct-fitting method described in Sarzi et
al. (2006).
This is similar to the method used by Sarzi et al. (2005) in the case
of the entire SUNNS sample, except in that the
spectral regions affected by nebular emission are no longer excluded
from the fitting process, since the emission-lines are treated as
Gaussian templates and fitted simultaneously with the stellar
templates to the observed spectra.
This has the advantage of maximising the spectral information
available to the fitting algorithm, which is crucial when dealing
with Sy2s where most, if not all, of the age- and
metallicity-sensitive stellar absorption features are heavily
contaminated by ionised-gas emission.

In fitting the spectra we also include reddening due to interstellar
dust, both in the Milky Way and in the sample galaxies, and due to
dust in the emission-line regions. The latter affects only the fluxes
of the emission-line templates, and is constrained by the expected
decrement of the Balmer lines.
In matching the nebular emission we imposed the same kinematics to all
forbidden lines, and likewise for the recombination lines.  Finally,
we allowed for complex line profiles using multiple Gaussian
components.

For the stellar population analysis it is critical to constrain the
strength of the high-order Balmer emission lines and the exact
profiles of the strongest lines. Accounting for their different
spectral resolution, we thus fitted both blue and red spectra at the
same time, which allows to exploit the longest wavelength leverage and
to match also the profiles of the strong \Ha\ and \Nii\ lines.

\placefigone
\placefigtwo

\section{Results}   
Using stellar population templates of solar metallicity ranging in age
from 1Myr to 10Gyr, we found that 10\% of the surveyed nuclei display
stellar populations of intermediate age, $\sim$1-2 Gyr old, whereas
the remaining Sy2s appear to be evenly split between objects
showing only very old stellar populations and nuclei requiring also an
additional blue featureless component, which here we characterise by
means of very young, few-Myr-old stars.
Figure~\ref{fig:fit} shows two examples of this last class of Sy2s,
which are also characterised by strong nebular emission.

Although the presence of a very young population seems required in
almost half of the sample, we need to keep in mind that a featureless
AGN continuum would probably fit these data as well.
In fact the relative dearth of intermediate-age stellar populations
seems to argue against interpreting such blue continua as due solely
to O-stars.
The short lifetime of such stars ($\sim3$Myr) would imply recurrent
and frequent star-formation episodes (every $\sim6$Myr while Seyfert
activity lasts) and the build-up of a detectable intermediate-age
stellar population if this behaviour have persisted for a 1 Gyr or
more.
Assuming that galactic nuclei cycle through different kind of active
phases and quiescent periods, based on the distribution of host
Hubble-types across our sample and on the fraction of Seyfert nuclei
in the nearby galaxy population (Ho et al. 1997) our sample nuclei
should have spent $\sim16\%$ of their recent history shining as Sy2s.
Over 1 Gyr they would have thus experienced $\sim27$ star-formation
episodes associated with Seyfert activity, leading to an
intermediate-age stellar population that should be detectable.
Such populations would indeed have luminosities comparable to that of
the blue continua we observe, since for a given mass and for standard
initial mass functions, a 1-Gyr-old stellar population is only 60 to
110 times dimmer than a few-Myr-old population.
%
%%
%At a given mass and for standard initial mass functions, a 1-Gyr-old
%stellar population is indeed only 60 to 110 times dimmer than a
%few-Myr-old population.

Furthermore, the luminosity of the blue component needed in our Sy2s
correlates well with the strength of the emission from highly-ionised
species such as \Nev$\lambda$3426 (Fig.~\ref{fig:Baldwin}), which
suggests that the featureless continuum arises from the central engine
rather than from star-forming regions.
Finally, in the Sy2s requiring a blue continuum, the characteristic
4650\AA\ ``bump'' due to Wolf-Rayet stars in strong and metal-rich
starbursts is (with one exception) either absent or can also be
explained by a superposition of forbidden lines such as
\Ariv$\lambda\lambda 4711,4740$, \Feiii$\lambda 4658$ and \Arv$\lambda
4625$. Forbidden lines from several other highly-ionised species are
also observed in these cases.

Given that the vast majority of our Sy2s do not show evidence of a BLR
even when zooming in the central $\sim$30 pc, these findings suggest
some of our targets could be ``true'' Type 2 sources, which are
predicted to form at low AGN luminosities by some models (e.g.,
Elitzur \& Shlosman 2006).
Alternatively, the small fraction of intermediate-age stellar
populations in our sample could be reconciled with on-going star
formation in roughly half of our objects if we consider the suggestion
of Nayakshin (2006) that only massive stars can form within few pc of
central SMBHs.
A population characterised by a top-heavy initial mass function would
indeed leave little trace of itself after a 1 or 2 Gyr. In this case,
however, such star-formation activity must be intimately connected to
the central accretion of material, in order to explain the correlation
of Fig.~\ref{fig:Baldwin}.

\begin{acknowledgments}
M. Sarzi is grateful to R. Cid-Fernandes for his useful suggestions. 
\end{acknowledgments}

\end{document}